\input amstex
\documentstyle{amsppt}

\hsize=4.75in
\vsize=8in
\NoBlackBoxes
\rightheadtext {Non Global Hyperbolicity and Locality} 
\leftheadtext {Bernard S. Kay} 
\topmatter 
\title Quantum Fields in Curved Spacetime: Non Global Hyperbolicity and
Locality 
\endtitle 
\author Bernard S. Kay
\endauthor 
\affil 
Department of Mathematics\break 
University of York\break
Heslington, York YO1 5DD, UK\break
(E-mail: bsk2\@york.ac.uk)
\endaffil 
\abstract
We briefly review the current status of the algebraic approach to 
quantum field theory on globally hyperbolic spacetimes, both axiomatic
-- for general field theories, and constructive -- for a linear
Klein-Gordon model. We recall the concept of F-locality, introduced in
the latter context in BS Kay: Rev. Math. Phys., Special Issue, 167-195
(1992) and explain how it can be formulated at an axiomatic level for a
general field theory (as a condition on algebras-with-net-structure) on
both globally hyperbolic and non globally hyperbolic spacetimes. We also
discuss the current status of the question whether/when algebras
satisfying F-locality can exist for the Klein-Gordon model on spacetimes
which are chronology violating.
\endabstract

\footnote{To be published in the proceedings of the conference `Operator
Algebras and Quantum Field Theory' held at Accademia Nazionale dei Lincei,
Rome, Italy, July 1996. (Editors S.~Doplicher, R.~Longo, J.~Roberts, 
L.~Zsido. Publisher [probably] International Press [distributed by the
AMS])}

\endtopmatter 
\document 
\heading
1. Introduction 
\endheading 

The extension of the algebraic approach to quantum field theory to the 
context of a general curved spacetime is of interest for (at least) two
reasons.  Firstly, quantum field theory in curved spacetime is by now
(see e.g. [1, 2]) a well-established theory for a wide range of
phenomena involving ``particle creation'' and ``vacuum polarization''
effects due to strong gravitational fields.  As long as one stays away
from scales where the quantum nature of the gravitational field itself
becomes important, this theory is expected to provide an excellent
approximate description\footnote{In cases where the back-reaction of
created particles etc. on the background geometry is expected to be of
importance, it is believed that this can be taken into account by
arranging that the background metric satisfies a semiclassical version
of Einstein's equation in which the right hand side is taken to be the
expectation value in a suitable state of the renormalized stress-energy
tensor operator of the relevant quantum fields.} for such phenomena, and
has led, amongst other things, to the remarkable prediction by Hawking
[3] of black-hole evaporation.   One of the features of the subject is
that one loses, in general, any single preferred quantum state which may
be regarded as a ``vacuum'' and the concept of ``particles'' becomes
vague and/or observer-dependent.  On the other hand, one is interested
in expectation values of local observables such as the field itself and
the stress-energy tensor which have a more objective existence.  The
algebraic approach to quantum field theory, in which the primary objects
are such local observables and the various states of interest may all 
be treated on an equal footing (as positive linear functionals on the
algebra of local observables) seems to be ideally suited (even
essential!) to discuss such  matters in a precise way,  and has been
used successfully e.g. in mathematical analyses of questions related to
the Hawking effect [4, 5].

Secondly, even if one's primary interest is in quantum field theory in
Minkowski space, one hopes that, by widening the context to a general
curved spacetime, one may achieve a deeper appreciation of the theory.
In particular, one hopes to attain a more completely {\it local}
understanding of many features of quantum field theory and to free
oneself from any unnecessary reliance on global features such as
Poincar\'e invariance.
 
The general context adopted in most discussions of quantum theory in
curved spacetime involves a quite thoroughgoing abandonment of many of 
the global features of Minkowski space.  It is usual to consider
spacetimes with different topologies  -- e.g. one might study quantum
fields on the  ``timelike cylinder'' -- i.e. the quotient of Minkowski
space by a fixed spacelike translation.  (The name refers to the, easily
visualized, two-dimensional version.)  Further, it is usual to consider
spacetimes which  have no isometries at all (other than the identity) --
e.g. one could obviously perturb the metric on the time-like cylinder so
as to remove both its time-translational and space-translational
symmetry.  However, there is one global feature which is usually
retained, namely {\it global hyperbolicity}.   Here we recall (see e.g.
[6, 7]) that a spacetime is said to be globally hyperbolic if it is
time-orientable and has a Cauchy surface -- i.e. an achronal spacelike
hypersurface which is intersected exactly once by every inextendible
causal curve.  (As a manifold, such a spacetime is then necessarily a
product of the Cauchy surface with the real line.)  There are of course
very good reasons for retaining this feature: Physically, realistic
spacetime models -- for the universe or for black holes etc. -- are
usually globally hyperbolic.  Mathematically, globally hyperbolic
spacetimes retain many of the qualitative features of the causal
structure of Minkowski space. Moreover, classical linear hyperbolic
equations such as the Klein-Gordon equation  
$$(\square - m^2)\phi=0\eqno{(1.1)}$$ 
(where $\square$ denotes the Laplace-Beltrami operator for the spacetime
metric, assumed here to have signature $(-+++)$) on such spacetimes
admit a well-posed Cauchy problem and have globally well defined
advanced and retarded (distributional) fundamental solutions
$\triangle^A$ and $\triangle^R$ satisfying 
$$(\square_x - m^2)\triangle^{A/R}(x,y)=\delta(x,y)$$ 
and supported, respectively, on the set of pairs $(x,y)$ for which $y$
is located to the future/past of $x$.  \footnote{Here we adopt the
customary loose practice of thinking of (bi)distributions as if they are
(bi)scalar functions.  We also identify functions with distributions by
integrating them with test functions using the natural volume element
provided by the (square root of the absolute value of the determinant of
the) metric.}

Nevertheless, it is of interest to contemplate the removal of this last
global condition and to ask whether it is possible (and how one might
try) to generalize the algebraic approach to quantum field theory to the
general class of not-necessarily globally hyperbolic spacetimes and, in
this short article, we shall briefly discuss some recent  and current
work by the author and collaborators on this problem [8, 9, 10, 11]. 
(For other work on related  questions, see e.g. [12].)   We shall be
particularly concerned with spacetimes which are chronology
violating\footnote{We shall, however, continue to assume that our
spacetimes are time-orientable.  For a discussion of why it is
difficult/impossible to quantize a field theory on a non-time-orientable
spacetime, see [8].  (See also [13].)}  (i.e. which have closed timelike
curves) a simple example of which is e.g. the {\it spacelike cylinder}
i.e. the quotient of Minkowski space by a fixed {\it timelike}
translation.  A subclass of chronology-violating spacetimes, namely
those with {\it compactly generated Cauchy horizons} (a simple
two-dimensional example is Misner space -- i.e. the quotient of [half
of] two-dimensional Minkowski space by a fixed Lorentz boost -- see e.g.
[6, 8] for details) may, according to the analysis in [14] be regarded
as models for spacetimes in which time-machines get manufactured and
thus our question is relevant to the currently topical question  [14,
15] of whether it is possible in principle to manufacture a time
machine.  

The basic geometrical consideration which underlies our subsequent
discussion has to do with the elementary result that, in any spacetime,
any point has neighbourhoods which are globally hyperbolic in their
intrinsic geometry. (We shall refer to them as {\it GH neighbourhoods}.)
Moreover, such GH neighbourhoods may be as small as we like in the sense
that any neighbourhood of any point contains a GH neighbourhood of that
point.  When the full spacetime is globally hyperbolic, one may always
find such a small-as-we-like GH neighbourhood around any point with the
property\footnote{Note however, that, even when the full spacetime is
globally hyperbolic, it is not true that every GH neighbourhood
has this property.  See e.g. the discussion of the ``helical strip''
example in [8].} that its intrinsic and induced causal structures
coincide (i.e. that pairs of points  which are timelike [respectively
spacelike, null] related in the intrinsic geometry of the neighbourhood
will also be timelike [respectively spacelike, null] related globally). 
In Minkowski space, examples of GH neighbourhoods with this property are
the familiar double cones.  However, in a non globally hyperbolic
spacetime, it may be impossible to find a GH neighbourhood of a given
point with equal intrinsic and induced causal structures. Thus, e.g. in
the spacelike cylinder, such neighbourhoods will clearly always contain
pairs of points which are intrinsically spacelike related but globally
timelike related. A key feature of spacetimes with compactly generated
Cauchy horizons is that (Proposition 2 of [9]) they necessarily have
some points  (the {\it base points} of [9]) in their Cauchy horizons
with the property
\proclaim{1.1 Property}
Every GH neighbourhood of $p$  contains a pair of points $(q,r)$ which
are intrinsically spacelike related but globally {\it null} related.
\endproclaim

\heading 
2. Nets of Local Algebras on Globally Hyperbolic Spacetimes
\endheading 

One expects a given quantum field theory on a given globally hyperbolic
spacetime $(M,g)$ to be describable by a C$^*$ algebra (with identity
$I$) with net structure, ${\Cal A}(M,g)$.  To say it has a net structure
means there is a preferred assignment of a subalgebra ${\Cal
A}(M,g)({\Cal O})$ to each open set $\Cal O$ with compact closure in $M$
-- satisfying the properties: 
\medbreak
\noindent
{\bf (1)\quad (Isotony.)} If ${\Cal O}_1$ is a subset of ${\Cal O}_2$,
then  ${\Cal A}(M,g)({\Cal O}_1)$ is a subalgebra of ${\Cal
A}(M,g)({\Cal O}_2)$. 
\smallbreak 
\noindent
{\bf (2)\quad (Spacelike Commutativity.)} If  ${\Cal O}_1$ and ${\Cal
O}_2$ are spacelike related, then ${\Cal A}(M,g)({\Cal O}_1)$ and ${\Cal
A}(M,g)({\Cal O}_2)$ commute. 
\medbreak
In one axiomatic approach [16] (which we shall adopt below ``by
default'') one expects the local algebras ${\Cal A}(M,g)({\Cal O})$ to
be W$^*$ algebras (specifically [hyperfinite type III$_1$] factors and
satisfying the condition of having trivial ``algebra at a point'') and
the total algebra to equal their C$^*$ inductive limit. In this
approach, the physically interesting states would be the locally normal
states. Alternatively, and essentially equivalently, one can adopt a
point of view where the local algebras are (smaller) $C^*$ algebras if
one imposes suitable conditions on the set of physically relevant states
[17, 18]. One advantage of the former approach is that the ${\Cal
A}(M,g)({\Cal O})$ are then expected [19, 5] to be large enough to
contain operators which would correspond to the (say exponentiated)
smeared (renormalized) stress-energy tensor operator.  
 
We remark that, given any subspacetime $(N,g)$ of $(M,g)$, one can
define the subalgebra (with its inherited net structure) ${\Cal
A}(M,g;N)$.  In our first mentioned (default) approach: 

\proclaim{2.1 Definition} 
${\Cal A}(M,g;N)$ is defined to be the C$^*$ completion of the
set of all ${\Cal A}(M,g)({\Cal O})$ for which the closure of $\Cal O$ is
contained in $N$.  (And, for such $\Cal O$, ${\Cal A}(M,g;N)({\Cal O})$
is defined to equal ${\Cal A}(M,g)({\Cal O})$.)
\endproclaim

In the case of a linear field theory [5], it is the structure referred
to in the alternative point of view above which is closer to what is
immediately constructable: Given the covariant Klein-Gordon equation
(1.1) on a globally hyperbolic spacetime $(M,g)$, there is a natural
[5, 20] ``minimal'' $C^*$ algebra ${\Cal B}(M,g)$ generated by
operators $W(F)$,  $F\in C_0^\infty(M)$ (formally related to the smeared
Hermitian quantum field $\phi(F)=\int_M\phi(x)F(x)|g|^{1\over 2}d^4x$ by
$W(F)=\exp(-i\phi(F))$) satisfying
$$W((\square-m^2)F)=0\eqno{(2.1)}$$ 
and
$$W(F_1)W(F_2)=\exp(-i\triangle(F_1,F_2)/2)W(F_1+F_2)\eqno(2.2)$$
where $\triangle=\triangle^A-\triangle^R$ is the advanced minus retarded
fundamental (antisymmetric distributional bi-)solution to Equation (1.1). This
algebra acquires a net structure by defining the local algebra ${\Cal
B}(M,g)({\Cal O})$ to be the closed linear span of the $W(F)$ for $F$
supported in ${\Cal O}$.
  
One then restricts attention to the set of states, $\omega$, whose
(unsmeared) symmetrized two-point functions have short-distance
singularities with the Hadamard form:
$$\omega(\phi(x)\phi(y)+\phi(y)\phi(x))=(1/2\pi^2)(u/\sigma+v\ln
|\sigma|+w)$$  
where $\sigma$ denotes the square of the interval between $x$ and $y$,
$u$ and $v$ are smooth two-point functions determined by the local
geometry and $w$ is a smooth two-point function which depends on the
state.  (Here, a suitable principle part prescription is understood in
giving this formula meaning for smeared fields -- see [5] for details.) 
We remark that (as conjectured in [21]) it is now known that this local
Hadamard condition on a two-point function (combined with the
restrictions on the two-point function due to positivity) prevents it
from having ``non-local spacelike singularities''.  This is one result
of a recent mathematical development [22] due to Radzikowski which,
importing concepts from microlocal analysis [23],  characterizes
Hadamard two-point functions in terms of their wave-front sets [22]. 
This development promises to have far-reaching applications beyond this
immediate problem.  (See also [9, 24, 25].)

It is also now known [18] that (as conjectured in [19]) the quasifree
Hadamard states are locally quasiequivalent.  One may thus enlarge each
of the local algebras ${\Cal B}(M,g)({\Cal O})$ to a well-defined W$^*$
algebra ${\Cal A}(M,g)({\Cal O})$ (by taking the double commutant of
each ${\Cal B}(M,g)({\Cal O})$ in the GNS representation of any
quasifree Hadamard state, and regarding the result as an abstract W$^*$
algebra) and define ${\Cal A}(M,g)$ to be the C$^*$ inductive limit of
all of these, thus obtaining a candidate algebra-with-net-structure in
the sense of the (default) axiomatic  approach discussed above.  We
refer to [18] (see also [26]) for the proof of this local
quasiequivalence result, and for partial results towards establishing
further expected properties of the resulting net of local algebras. 

Before turning to discuss whether/how one might generalize the above
axioms/constructions to non globally hyperbolic spacetimes, we draw
attention to a locality property (pointed out in [8]) inherent in the
construction of   ${\Cal B}(M,g)$ which will be of relevance to that
discussion:  Namely, it possesses:

\proclaim{2.2 The F-Locality Property (Klein-Gordon Version)} Any               
neighbourhood of any point $p$ in $(M,g)$ contains a GH neighbourhood $N$       
of $p$ such that the map which sends the element denoted ``$W(F)$'' in          
${\Cal B}(M,g)(N)$ to the element denoted ``$W(F)$'' in ${\Cal B}(N,g)$         
(where $F$ ranges over $C_0^\infty(N)$) extends to an isomorphism.              
\endproclaim

(Here, we make the obvious adaptation from the discussion in [8] which was 
couched in terms of a slightly different technical framework based 
on $*$ algebras of smeared fields.)

To see that this property holds, choose $N$ to have the same induced and
intrinsic causal structure (see Section 1).  The isomorphism is then
clear since the advanced minus retarded fundamental solution for
$(M,g)$, when restricted to $N$, will coincide with the intrinsic
advanced minus retarded fundamental solution for $(N,g)$.

We remark here that the algebra-with-net-structure ${\Cal A}(M,g)$
constructed, as sketched above, for the Klein-Gordon model will inherit
an obvious version of this property.  This may be best expressed by
slightly changing viewpoint and thinking of the resulting property as a
property of the map
$$(M,g)\mapsto {\Cal A}(M,g)$$
from the set of all globally hyperbolic spacetimes to the set of
algebras-with-net-structure:

\proclaim{2.3 The F-Locality Property (Generalizable Formulation)} 
Given any globally hyperbolic spacetime $(M,g)$, every neighbourhood of
every point $p\in M$ contains a GH neighbourhood $N$ of $p$ such that
there is a net-structure-preserving isomorphism between ${\Cal
A}(M,g;N)$ and ${\Cal A}(N,g)$.  (Here, ${\Cal A}(M,g;N)$ is defined as
in Definition 2.1.) 
\endproclaim

It seems reasonable to expect that (as we indicate by the above name) 
this property should hold quite generally -- i.e. for the (default)
algebras-with-net-structure of not-necessarily-linear quantum field
theories on globally hyperbolic spacetimes.

\heading
3. The F-Locality Condition
\endheading

In attempting to extend the algebra construction(s), sketched above for
the Klein-Gordon field, to the case of a general non globally hyperbolic
spacetime, one immediately encounters the difficulty that there may be
no globally defined advanced-minus-retarded fundamental solution and
thus no obvious replacement for the relations (2.2).  (As far as axioms
are concerned, notice e.g. that, on many spacetimes -- the spacelike 
cylinder being one example -- there are no spacelike related pairs of
points and the Spacelike Commutativity Axiom would, in consequence, be
empty of content.)  The basic idea, suggested in [8], for overcoming
this problem is to replace the global relation (2.2) by a suitable local
remnant. There is considerable leeway [8] as to how this might be
achieved, but one simple proposal (for a necessary condition) is to
demand that the relations (2.2) continue to hold on ``sufficiently small
finite neighbourhoods'' of each point.  More precisely, one demands of
any candidate replacement ${\Cal B}(M,g)$ for the minimal C$^*$ algebra
that it should still  contain elements $W(F)$, $F\in C_0^\infty(M)$
which (whatever other relations they may satisfy) must satisfy, in
addition to the relations  (2.1) above, the {\it F-Locality Condition}
(cf. the above F-Locality Property): Every point $p\in M$ should have a
GH neighbourhood $N$ such that the $W(F)$, $F\in C_0^\infty(N)$ satisfy
Relation (2.2) above with $\triangle$ taken to be the intrinsic advanced
minus retarded fundamental solution on $(N,g)$. In other words,
defining, for any open set $\Cal O$, the local algebra  ${\Cal
B}(M,g)({\Cal O})$ to be the closed linear span of the $W(F)$ for
$F$ supported in $\Cal O$:

\proclaim{3.1 The F-Locality Condition (Klein-Gordon Version)} Every
point $p\in M$ should have a GH neighbourhood $N$ such that the
map which sends the element denoted ``$W(F)$'' in ${\Cal B}(M,g)(N)$ to
the element denoted ``$W(F)$'' in ${\Cal B}(N,g)$ (where $F$ ranges over
$C_0^\infty(N)$) extends to an isomorphism.  \endproclaim
  
We remark here that one could generalize this proposal to a general
axiomatic framework by assuming we are given a map  
$$(N,g)\mapsto {\Cal A}(N,g)$$ 
from globally hyperbolic spacetimes to (default)
algebras-with-net-structure and demanding, for any proposed 
algebra-with-net-structure ${\Cal A}(M,g)$ on a given non globally hyperbolic
spacetime $(M,g)$ that it satisfy (cf. Property 2.3 above):

\proclaim{3.2 The F-Locality Condition (Generalizable Formulation)} 
Every point $p\in M$ must have a GH neighbourhood $N$  such that there
is a net-structure-preserving isomorphism between  ${\Cal A}(M,g;N)$
(defined as in Definition 2.1) and ${\Cal A}(N,g)$. 
\endproclaim

\heading
4. Difficulties with Chronology-Violating Spacetimes
\endheading

Returning to the framework discussed above for the Klein-Gordon
equation, and focussing now on chronology violating spacetimes (see
Section 1) we discuss what is now known about whether/when (quite apart
from what might be a full set of necessary and sufficient conditions)
there can exist any algebras ${\Cal B}(M,g)$ at all which satisfy the
F-Locality Condition stated above.

For a spacetime $(M,g)$ with compactly generated Cauchy horizon, it is
now known quite generally [9] that, with the mild technical assumption
that the algebra admits a state $\omega$ in which the non-exponentiated
smeared two-point function $\omega(\phi(F_1)\phi(F_2))$
exists\footnote{Here, we are making the obvious adaptation of  Theorem
1$'$ of Reference [9] (which like Reference [8] is couched in the
slightly different technical framework of $*$ algebras of smeared
fields) to the present framework.}  and is distributional, then there is
no algebra ${\Cal B}(M,g)$ which satisfies F-locality. In fact, this
no-go result must hold in any spacetime which contains a point $p$
satisfying Property 1.1.  (Or the more general property which results if
one replaces the word ``spacelike'' there by ``non-null''.) The proof
relies on a micro-local version ([23] Volume IV) of the Propagation of
Singularities Theorem applied to distributional bisolutions to the
Klein-Gordon equation.   Roughly speaking, this tells us that a
bisolution which is singular for pairs of points $(q,q')$ where $q$ is
fixed, say, and $q'$ ranges over a portion of a null geodesic, must
remain singular when $q'$ is allowed to range over the full null
geodesic.  (In the theorem, what actually propagates [along
bicharacteristics] is the ``wave front set'' which consists of [pairs
of] points in the cotangent space to $M$ representing pairs of
``points-together-with-codirections'' at which the distributional
bisolution fails to be smooth.  See [23] Volume IV.) One focusses on the
quantity $\omega(\phi(F_1)\phi(F_2)-\phi(F_2)\phi(F_1))$.  Regarded as
an (antisymmetric) distributional bisolution to the Klein-Gordon
equation, if one assumes that F-locality holds, this must clearly
coincide, in some GH neighbourhood $(N,g)$ of $p$ with $i$ times the
intrinsic advanced minus retarded fundamental (antisymmetric
distributional bi-)solution $\triangle_{(N,g)}$.   It is well-known that
such a fundamental solution will be singular for nearby intrinsically
null related pairs of points in $(N,g)$ but smooth for intrinsically non
null related such pairs (in fact it is zero if they are spacelike
related!) and, in particular, it will be smooth (zero) for the pair
$(q,r)$ of points which we know $N$ must contain by Property 1.1  which
are intrinsically non null related but globally {\it null} related. 
However, since this pair is  globally null related, we know that we can
propagate the singularity from pairs $(q,q')$  -- where $q'$ is close to
$q$ and ranges over a portion of the null  geodesic joining $q$ to $r$
-- to the pair $(q,r)$ thus obtaining a contradiction. (See [9] for a
full discussion and for other related theorems which, quite
independently from any question as to what conditions a field algebra
should satisfy, rule out the existence of any everywhere [``weakly''
[9]] locally Hadamard distributional bisolution on $(M,g)$ and imply
that expectation values in Hadamard states [defined in the ``initial
globally hyperbolic  region''] of the renormalized stress-energy tensor
must necessarily become singular at any base point. See also [10].)

Turning to consider chronology-violating spacetimes which do not contain
any points satisfying Property 1.1, the situation for F-locality seems
to be rather delicate.  In fact, it was pointed out in [8] that, in the
case that $(M,g)$ is the (four-dimensional)  spacelike cylinder (see
Section 1) and one specializes to the {\it massless} Klein-Gordon
equation, one {\it can} construct an F-local algebra ${\Cal B}(M,g)$.
(In the language of [8], for this field theory model, this spacetime is
{\it F-quantum compatible}):   One simply follows the usual construction
(see Section 2) used in the globally hyperbolic case, replacing
$\triangle$ in Equation (2.2) by the result of ``wrapping'' the
Minkowski space fundamental solution around the cylinder.   (i.e.
thinking of the spacetime as consisting of equivalence classes $[x]$ of
points in Minkowski space where ``$x$ is equivalent to $y$'' means that,
in some inertial coordinate system, their space coordinates coincide,
while their time coordinates differ by multiples of some fixed time
interval, and with a notation that treats distributions as if they are 
functions, one takes 
$$\triangle([x],[y])=\sum_{y\in [y]}\triangle_{\hbox{Minkowski}}(x,y).$$
One easily sees that this produces an  F-local theory because, in four
dimensions, the fundamental solution to the Klein-Gordon equation
vanishes for timelike related as well as for spacelike related pairs of
points.  It is also not difficult to convince oneself that the
construction can be modified to cope with the massless wave equation on
the two-dimensional spacelike cylinder.  However, it was left as an open
question in [8] whether any F-local algebra exists on the spacelike
cylinder in the case of the massive Klein-Gordon  equation.  Clearly,
the above wrapping procedure will now fail because the massive Minkowski
fundamental solution does not vanish for timelike related pairs of
points. (Notice however that it is {\it smooth} for such pairs and, in
consequence, the wrapping procedure will give a theory which satisfies
the weaker condition of {\it F-locality modulo $C^\infty$} as defined
in Condition 4.1 below.)

It was shown by Fewster and Higuchi [27] that, perhaps surprisingly, one
can nevertheless construct F-local algebras on the spacelike cylinder in
this massive Klein-Gordon case.  The constructions of [27] rely heavily,
however, on the translational invariance of the spacelike cylinder and
it is thus not clear from this work to what extent this result is
``generic''.  In particular, one can ask (for a given field theory)
whether the property of being F-quantum compatible is stable under
perturbations in the metric. Obviously, in the case of a (conformally
coupled) massless model, one will have stability of any F-quantum
compatible spacetime in this sense under conformally flat perturbations.
In some F-quantum compatible examples (such as conformally coupled
massless fields on compactified Minkowski space or on the
two-dimensional null strip --  see [9]) arbitrarily small non-conformal
perturbations of the metric will, however, lead to the existence of
points satisfying Property 1.1 and hence these models will obviously be
unstable under unrestricted perturbations by the no-go proof of [9] as
sketched above.  On the other hand, sufficiently small perturbations of
the metric on the spacelike cylinder will not lead to the existence of
such points and a separate analysis is required:   The next simplest
case to examine is perhaps that of massive fields on the spacelike
cylinder under conformal perturbations of the metric.  In recent work by
Fewster, Higuchi and the author [11], we have found that, in two
dimensions, there is a large class of one-parameter families of such 
perturbations of the spacelike cylinder under which its F-quantum
compatibility for the massive Klein-Gordon model is unstable.  (See also
[28].) However, it has proven more difficult to establish such an
instability result in higher dimensions.  We have, however, now found
[11] a small number of one-parameter families of conformal perturbations
of the spacelike cylinder -- by perturbations which respect its space
translational invariance but break its time translational invariance --
for which one can show that (for arbitrarily small values of the
parameter) there does not exist any {\it space-translationally
invariant} F-local field algebra for the massive Klein Gordon equation.

One possible conclusion to all of this is that, ``generically'',
F-locality rules out essentially {\it all} chronology violating
spacetimes and one might be tempted to draw the moral that imposing the
usual ``globally hyperbolic'' rules in the small more or less enforces
global hyperbolicity in the large! (Note also/cf. the way in which
non-time orientable spacetimes are ruled out in [8, 13].)  Notice
however that this conclusion could be different if, in making precise
what one means by ``the usual rules in the small'',  one were to adopt a
weaker notion than F-locality.  We remark in this connection that one
could e.g. replace F-locality by  the property of {\it F-locality modulo
$C^\infty$} 

\proclaim{4.1 The F-Locality Modulo $C^\infty$ Condition (Klein-Gordon
Version)}  Every point $p\in M$ should have a GH neighbourhood $N$ such
that ${\Cal B}(M,g)(N)$ is isomorphic  (with ``$W(F)$'' mapping to
``$W(F)$'') to some modified ${\Cal B}(N,g)$, where one is allowed to
change the rules for quantizing on $(N,g)$ by replacing $\triangle$
(i.e. the advanced minus retarded fundamental solution on $(N,g)$) in
Equation (2.2) by $\triangle + F$ where $F$ is a smooth antisymmetric
bisolution to the Klein-Gordon equation on $(N,g)$. \endproclaim  

It may be interesting to ask whether/how one can provide a
``generalizable formulation'' of this condition in the spirit of
Property 2.3 and Condition 3.2.

It is easy to see that this relaxation of the F-locality condition would
not help in the case of spacetimes with compactly generated Cauchy
horizons (or of any spacetime containing a point $p$ for which Property
1.1 [generalized as above] holds) since the proof (sketched above) of
the no-go theorem of [9] will clearly still go through.  But, it is easy
to see that small perturbations of the spacelike cylinder will continue
to admit algebras ${\Cal B}(M,g)$ satisfying this condition even for the
massive Klein-Gordon equation.  To see this, it suffices to  unwrap the
spacetime to obtain a periodic perturbation of Minkowski space. As long
as this is globally hyperbolic, one can consider its advanced minus
retarded fundamental solution $\triangle_{\hbox{periodic}}$ and 
construct ${\Cal B}(M,g)$ by following the rules for globally hyperbolic
spacetimes, taking $\triangle$ in Equation (2.2) to be the result of
wrapping (in the sense discussed above)  $\triangle_{\hbox{periodic}}$
around the spacetime. Note however that this construction would violate
the axiom of spacelike commutativity even in arbitrarily small (GH)
neighbourhoods.

In conclusion, one can ask:  What do we learn from all this concerning
the physically possible states of the world?  Our tentative answer would
be: If we restrict our attention to those states which admit a
description in terms of a classical spacetime with a net of local
algebras, and insofar as such nets satisfy similar ``laws in the small''
to the familiar ``globally hyperbolic'' laws, then {\it either}
chronology violating spacetimes are ruled out, {\it or} if they can be
realized physically, then one would in principle be able to detect this
fact locally by observing local violations of spacelike commutativity. 
(In the case of spacetimes with compactly generated Cauchy horizons, or
more generally of any spacetime containing a point $p$ for which
Property 1.1 [generalized as above] holds, the further no-go theorems of
[9] which concern the singularity in the stress-energy tensor rule such
spacetimes out also in the sense that they cannot arise as semiclassical
solutions to Einstein's equations.) 

\heading
5. Acknowledgements
\endheading

It is a pleasure to acknowledge the contributions made by my
collaborators Claes Cramer, Marek Radzikowski and Robert Wald
to the recent joint work described in this article.  Also, and
especially, I thank Atsushi Higuchi and Christopher Fewster for
generously permitting me to describe some of our current joint work here
prior to its publication.

The work described here was supported in part by EPSRC grant GR/K/29937
to the University of York.

\heading 
References 
\endheading

\item{[1]} N.D.~Birrell and P.C.W.~Davies: {\it Quantum Fields in Curved
Space}, Cambridge University Press, Cambridge (1982).

\item{[2]} R.M.~Wald: {\it Quantum Field Theory in Curved Spacetime and
    Black Hole Thermodynamics}, Chicago University Press, Chicago (1994).

\item{[3]} S.W.~Hawking: {\it Particle creation by black holes}, 
  Commun.~Math.~Phys. {\bf 43} (1975) 199-220.
  
\item{[4]} K.~Fredenhagen and R.~Haag: {\it On the derivation of Hawking
radiation associated with the formation of a black hole}, Commun.~Math.~Phys.
{\bf 127} (1990) 207-284.

\item{[5]} B.S.~Kay and R.M. Wald: {\it Theorems on the uniqueness and
    thermal properties of stationary, nonsingular, quasifree states on
    spacetimes with a bifurcate Killing horizon}, Physics Reports {\bf
    207} (1991) 49-136.  See also:\hfil\break 
    B.S. Kay: {\it Sufficient conditions for quasifree states and an
      improved uniqueness theorem for quantum fields on spacetimes with
      horizons}, J.~Math.~Phys. {\bf 34} (1993) 4519-4539.
    
\item{[6]} S.W.~Hawking and G.F.R.~Ellis: {\it The Large Scale Structure of
    Space-Time}, Cambridge University Press, Cambridge (1973).

\item{[7]} R.M.~Wald: {\it General Relativity}, Chicago University
  Press, Chicago (1984).

\item{[8]} B.S.~Kay: {\it The principle of locality and quantum field
    theory on (non globally hyperbolic) curved spacetimes}, Rev.~Math.~Phys.
  {\bf Special issue dedicated to R. Haag} (1992) 167-195.

\item{[9]} B.S.~Kay, M.J.~Radzikowski and R.M.~Wald: {\it Quantum field
    theory on spacetimes with a compactly generated Cauchy horizon}, 
  Commun.~Math.~Phys. (in press, 1997).
  
\item{[10]} C.R.~Cramer, B.S.~Kay: {\it Stress-energy must be singular
    on the Misner horizon even for automorphic fields}, Class.~Quantum~Grav.
  {\bf 13} (1996) L143-L149. 

\item{[11]} C.J.~Fewster, A.~Higuchi and B.S.~Kay: {\it How generic is
    F-locality? Examples and counterexamples}.  (In preparation.)

\item{[12]} U.~Yurtsever: {\it Algebraic approach to quantum field theory on
  non-globally-hyperbolic spacetimes}, Class.~Quantum~Grav. {\bf 11}
(1994) 999-1012 

\item{[13]} J.L.~Friedman and A.~Higuchi: {\it Quantum field theory in
Lorentzian universes-from-nothing}, Phys.~Rev. {\bf D52} (1995)
5687-5697. 

\item{[14]} S.W.~Hawking: {\it The chronology protection conjecture},
Phys.~Rev. {\bf D46} (1992) 603-611.

\item{[15]} K.S.~Thorne: {\it Closed timelike curves}.  In R.J.~Gleiser,
    C.N. Kozameh and O.M. Moreschi (editors): {\it Proceedings of the 13th 
    International Conference on General Relativity and Gravitation. (Cordoba,
   Argentina 1992)}, (pages 295-315) Institute of Physics Publishing,
Bristol (1993).
    
\item{[16]} R.~Haag, H.~Narnhofer and U.~Stein: {On quantum field theory
    in gravitational background}, Commun.~Math.~Phys. {\bf 94} (1984) 219-238.

\item{[17]} R. Haag: {\it Local Quantum Physics}, Springer Verlag, Berlin
  Heidelberg New York, (1992).

\item{[18]} R. Verch: {\it Local definiteness, primarity and
    quasiequivalence of quasifree Hadamard states in curved spacetime},
  Commun.~Math.~Phys. {\bf 160} (1994) 507-536.

\item{[19]} B.S.~Kay: Talk at {\it 10th International Conference on
General Relativity and Gravitation (Padova, 1983)}, reported in Workshop
Chairman's Report by A.~Ashtekar in the Proceedings (pages 453-456):
B.~Bertotti et al. (editors) Reidel, Dordrecht (1984).  

\item{[20]} J.~Dimock: {\it Algebras of local observables on a manifold},
  Commun.~Math.~Phys. {\bf 77} (1980) 219-228.
  
\item{[21]} B.S.~Kay: {\it Quantum field theory in curved spacetime}. 
In: K.~Bleuler and M.~Werner (eds.) {\it Differential Geometrical
Methods in Theoretical Physics}, Reidel, Dordrecht (1988) (pages 373-393)
and\hfil\break 
G.~Gonnella and B.S.~Kay: {\it Can locally Hadamard
quantum states have non-local singularities?} Class. Quantum Grav. {\bf
6} (1989) 1445-1454.

\item{[22]} M.J.~Radzikowski: {\it The Hadamard Condition and Kay's
Conjecture in (axiomatic) Quantum Field Theory on Curved Space-time}
Ph.D. dissertation, Princeton University, (1992).  Available through
University Microfilms International, 300 N.~Zeeb Road, Ann Arbor,
Michigan 48106 U.S.A.;\hfil\break
M.J.~Radzikowski: {\it Micro-local approach to the Hadamard condition
in quantum field theory on curved space-time}, Commun.~Math.~Phys.
{\bf 179} (1996) 529-553;\hfil\break
M.J.~Radzikowski: {\it A local-to-global singularity theorem for quantum
field theory on curved space-time}, Commun.~Math.~Phys. {\bf 180} (1996)
1-22. (See also the Appendix by R.~Verch.)

\item{[23]} L.~H\"ormander: {\it The Analysis of Linear Partial Differential
    Operators}  Vols. I-IV, Springer-Verlag, Berlin Heidelberg New York 
    (1983-1986).

\item{[24]} R.~Brunetti, K.~Fredenhagen and M.~K\"ohler: {\it The
    microlocal spectrum condition and Wick polynomials of free fields on curved
    spacetimes}, Commun.~Math.~Phys. {\bf 180} (1996) 633-652.

\item{[25]} R.~Brunetti and K.~Fredenhagen: {\it Interacting quantum fields
    in curved space: Renormalizability of $\varphi^4$} (gr-qc 9701048,
   and these proceedings).

\item{[26]} C.~L\"uders and J.E.~Roberts: {\it Local quasiequivalence
    and adiabatic vacuum states}, Commun.~Math.~Phys. {\bf 134} (1990) 29-34.

\item{[27]} C.J.~Fewster and A.~Higuchi: {\it Quantum field theory on certain
    non-globally hyperbolic spacetimes}, Class. Quantum Grav. {\bf 13} (1996)
  51-61.

\item{[28]} C.J.~Fewster: {\it Bisolutions to the Klein-Gordon equation
    on two-dimensional cylinder spacetimes}.  (In preparation.)

\enddocument
\bye